\documentclass[
 twocolumn,secnumarabic,amssymb,
amsmath,nofootinbib, aps, prl, showpacs]{revtex4}

\begin{document}

\title{The linear Ising model and its
analytic continuation, random walk}

\author{B. H. Lavenda}
\email{bernard.lavenda@unicam.it}
\affiliation{Universit\'a degli Studi, Camerino 62032 (MC) Italy}
\date{\today}

\newcommand{\half}{\mbox{\small$\frac{1}{2}$}}
\newcommand{\third}{\mbox{\small$\frac{1}{3}$}}
\newcommand{\twothirds}{\mbox{\small$\frac{2}{3}$}}

\newcommand{\fourth}{\mbox{\small$\frac{1}{4}$}}
\newcommand{\fourthirds}{\mbox{\small$\frac{4}{3}$}}
\newcommand{\fivethirds}{\mbox{\small$\frac{5}{3}$}}
\newcommand{\threehalves}{\mbox{\small$\frac{3}{2}$}}
\newcommand{\summ}{\sum_{i=1}^m\,}
\newcommand{\sumj}{\sum_{j=1}^k\,}
\newcommand{\suml}{\sum_{\ell=1}^m}
\newcommand{\sumll}{\sum_{\ell=k+1}^m}
\newcommand{\ebar}{\bar{E}}
\newcommand{\sumn}{\sum_{i=1}^n\,}
\newcommand{\onem}{\mbox{\small$\frac{1}{m}$}}
\newcommand{\onen}{\mbox{\small$\frac{1}{n}$}}
\newcommand{\mtwo}{\mbox{\small$\frac{m}{2}$}}
\newcommand{\ntwo}{\mbox{\small$\frac{n}{2}$}}
\newcommand{\onenone}{\mbox{\small$\frac{1}{n-1}$}}
\newcommand{\sumntwo}{\sum_{i=2}^n\,}

\begin{abstract}
A generalization of Gauss's principle is used to derive the error laws  
corresponding to Types II and
VII distributions in Pearson's classification scheme. 
Student's $r$-pdf  (Type II) governs the distribution of the internal energy
of a uniform, linear chain, Ising model, while 
analytic continuation of the uniform
exchange energy  converts it into a Student $t$-density (Type VII) 
for the position of a
random walk in a single spatial dimension. Higher 
dimensional spaces, corresponding to larger degrees of freedom and generalizations to
multidimensional Student $r$- and $t$-densities, are obtained
by considering independent and identically distributed
random variables, having rotationally invariant densities, whose entropies are
additive and generating functions are multiplicative.
\end{abstract}
\maketitle
A relation between the Ising problem and random walks has been conjectured a number of
times over the years \cite{Barber}. Rather than one of technique, we shall show in this
letter that the two are inter convertible through analytic continuation of the 
uniform exchange energy which transforms a probability density function (pdf) of finite
range into one of infinite range. Specifically, a Student $r$-pdf \cite{Korn}, of limited
range, which governs the distribution of the internal energy of a linear, uniform
Ising chain, is transformed into  Student's $t$-pdf, of unlimited range, for
the position of a random walker along an infinite linear chain, through analytic 
continuation of the uniform exchange energy between spins.
These two families of distributions will be derived as error laws from a generalization of Gauss's 
principle, which will serve in uniting these  two seeming unrelated phenomena. 
\par
Gauss's principle asserts that if the probability of an error, given the true value, is a
function of the error alone, and the log-likelihood function is maximum when the true
value coincides with the mean of the observed values, then the law of error must be 
normal \cite{Jeffreys}. 
However, many other error laws exist, such as the beta pdf, in which the 
probability of an error is not a function of the error 
alone \cite[pp. 165]{Jeffreys}. Moreover, other error laws  tend to the normal law only 
in the asymptotic limit, where the number of observations increases without limit. 
 Student's $r$- and $t$-pdfs  fall into this category.  \par
A law of error for determining the probability of deviations from a location parameter
$\lambda$ is given in the form:
\begin{equation}
P(dx|\lambda)=p(x-\lambda)\,dx.\label{eq:Pr}
\end{equation}
The method of maximum likelihood switches the roles of variate, $x$, and
parameter, $\lambda$, thereby converting the probability (\ref{eq:Pr}) into a likelihood
function. Maximum likelihood is determined from a solution of
\begin{eqnarray}
\lefteqn{\frac{d}{d\lambda}\log\mathcal{L}(\lambda)}\nonumber\\
&:= & \frac{d}{d\lambda}\log\prod_{i=1}^{n}p(x_i-\lambda)=\sum_{i=1}^{n}
\frac{p^{\prime}(x_i-\lambda)}
{p(x_i-\lambda)}=0,\label{eq:lle}
\end{eqnarray}
where the prime stands for differentiation with respect to $\lambda$. 
\par
If the arithmetic mean,
\begin{equation}\sum_{i=1}^{n}(x_i-\lambda)=0,\label{eq:arithmetic}
\end{equation}
is the most probable value of the quantity measured,
then the normal law follows by integration of the log-likelihood equation. The 
log-likelihood equation is formed by setting deviations from (\ref{eq:lle}) 
proportional to deviations
from (\ref{eq:arithmetic}), where the constant of proportionality is conveniently 
taken as the second derivative of an, at least, twice differentiable continuous 
function \cite{Keynes}.
The normal law corresponds to the case of a negative unit weight
\begin{equation}
\frac{p^{\prime}(x_i-\lambda)}{(x_i-\lambda)p(x_i-\lambda)}=\sigma_i. \label{eq:weight}
\end{equation}
Integrating and using
the integration function to complete the square yields the normal pdf.
For any other constant weight, we would get the weighted average 
\cite[pp. 214]{Jeffreys}
\[\sum_{i=1}^{n}\sigma_i(x_i-\lambda)=0,\]
with $\sumn\sigma_i=1$, rather than the arithmetic mean, (\ref{eq:arithmetic}).\par 
Actually, (\ref{eq:weight}) must be true for whatever 
$i$ is chosen. Hence, it must be independent of $i$, and, at most, can be a function
of the location parameter $\lambda$. Except for the normal law, where $\sigma$
is a negative constant, the Poisson law, where $\sigma=-\lambda^{-1}$, the binomial, and 
negative, binomial laws, where $\sigma_{\mp}=-[\lambda(1\mp\lambda)]^{-1}$,
respectively, other error laws cannot be derived from the conventional form
of Gauss's principle \cite{Lavenda91}. 
One example is  Student's $t$-pdf,
even though it tends to the normal law as the number of degrees of freedom 
increases without limit. The $t$-pdf is nearly normal at
the center but it has fatter tails than the normal pdf because  higher order moments
do not exist. Another example 
is Student's $r$-pdf, notwithstanding the fact that all the moments of the 
distribution exist. \par
We \cite{Lavenda91} have identified the weights, $\sigma_i$, as the second 
derivatives of the entropy. The weights we will consider here 
\cite[pp. 210--211]{Jeffreys},
\begin{equation}
\sigma_{\mp}=-\left(1\mp \frac{(x-\lambda)^2}{s^2}\right)^{-1}, \label{eq:sigma}
\end{equation}
will lead to Pearson distributions of Types II (negative sign) and VII (positive sign). 
We assume that the information regarding the values of the location, $\lambda$, and
scaling, $s$, parameters depend on
\cite[p. 379]{Jeffreys}
\[
u=\frac{x-\lambda}{s} \]
alone. \par
In the case of Type II distributions a single integration of $\sigma_{-}$ gives the
\lq equation of state\rq
\begin{equation}
\beta=-\tanh^{-1}u, \label{eq:beta}
\end{equation}
where we will soon appreciate $\beta$ as the intensive variable conjugate to the extensive
variable $u$. In essence, (\ref{eq:beta}) contains the statement of the
 second law of thermodynamics
\[\frac{\partial S}{\partial u}=\beta,\]
where $S(u)$ is the entropy. A
further integration yields the entropy reduction
\begin{eqnarray}
\lefteqn{\Delta S(u):=S(u)-S(0)}\label{eq:Ising}\\
& = &-u\tanh^{-1}u-\half\log\left(1-
u^2\right)+\mbox{const.}, \nonumber
\end{eqnarray}
where $S(0)$ is the maximum entropy in the \lq high temperature limit\rq\ 
$\beta\rightarrow0$, where $u=0$.\par 
Introducing the equation of state (\ref{eq:beta}) into
(\ref{eq:Ising}) gives
\begin{equation}
\Delta S=-\beta\tanh\beta+\log\cosh\beta+\mbox{const.}
\label{eq:S-Ising}
\end{equation}
 Setting $\beta=J/T$, where $J$ is the uniform
exchange energy between spins, and $T$ the absolute temperature in energy units where
Boltzmann's constant is unity, $u$ in (\ref{eq:beta}) is identified as the ratio of 
the internal energy per unit
spin to the exchange energy, and (\ref{eq:S-Ising}) is the zero-field entropy reduction per spin of
the uniform, linear chain, Ising model \cite{Wannier}. Since there are $2^m$ configurations, the maximum
entropy per spin is $S(0)=\log 2$.\par
 For $m$ spins (\ref{eq:Ising}) will be $m$ times
as large. Viewed as the Legendre transform
\begin{equation}
\Delta S(u)-\frac{\partial\Delta S}{\partial u}u=\log\mathcal{Z}(\beta), 
\label{eq:Legendre}
\end{equation}
 the generating function per spin is identified as
\begin{equation} 
\mathcal{Z}(\beta)=\cosh\beta. \label{eq:Z-Ising}
\end{equation}
The generating function for $m$ spins  is \cite{Stanley}
\begin{equation}
\mathcal{Z}_m(\beta)=
\cosh^{m-1}\beta, \label{eq:Z-Ising-m}
\end{equation}
which is the hallmark of an infinitely divisible distribution \cite{Finetti}.\par
The Legendre transform (\ref{eq:Legendre}) is asymptotically equivalent to the Laplace
transform
\begin{equation}
\mathcal{Z}_m(\beta)=\int_{-1}^{0}\,e^{-m(\beta u-\Delta S(u))}\,du
\label{eq:Laplace}
\end{equation}
in the limit as $m\rightarrow\infty$. The function, $\beta u-\Delta S(u)$, has 
a minimum on $[-1,0]$  at the interior point (\ref{eq:beta}). Employing
Laplace's method, the integral (\ref{eq:Laplace}) can be evaluated by developing the
exponent in a Taylor series to second order about the minimum (\ref{eq:beta}), and using
the fact that $m$ is so large that the value of the resulting Gaussian integral is 
barely changed when the lower limit of integration is set equal to $-\infty$. We then 
obtain
\[
\mathcal{Z}_m(\beta)=\sqrt{\frac{\pi}{2m}}\cosh^{m-1}\beta \]
for the expression of the generating function of $m$ spins. Consequently, 
(\ref{eq:Z-Ising})
and (\ref{eq:Z-Ising-m}) are logarithmically equivalent.
\par
The entropy reduction determines the univariate pdf through Boltzmann's principle 
\cite{Lavenda85}:
\begin{equation}
p(u)=K\; e^{\Delta S(u)}, \label{eq:Boltzmann}
\end{equation}where $K$ is a constant. Introducing 
(\ref{eq:Ising}) into (\ref{eq:Boltzmann}) gives
\begin{equation}
p(u)=K\frac{e^{-u\tanh^{-1}u}}{(1-u^2)^{\half}}=
K\left(1-u^2\right)^{\half}+O\left(u^4\right).\label{eq:p-single}
\end{equation}
\par
 Consider a uniform distribution of mass inside 
a hypersphere of $m$-dimensions with unit radius. The energies 
$U_1,\ldots,U_{m}$ of 
$m$ spins are supposed to be independent and identically distributed (iid), and, 
in addition, they have rotationally invariant densities. Such a hypersphere will 
be spanned by the coordinates, $u_1,\ldots,u_{m}$. The Euclidean distance from 
the origin to any point 
\[|u|=\left(u_1^2+\cdots+u_{m}^2\right)^{\half}.\]
The section at $|u|$ has radius $\sqrt{1-u^2}$, and
an \lq area\rq\ proportional to $(1-u^2)^{\half(m-1)}$. 
Hence, it has a density 
\begin{eqnarray}
p_{m}(u) & = & K e^{(m-1)\Delta S(u)}\label{eq:p-Ising}\\
& \simeq & K
\left(1-u^2\right)^{\half(m-1)}+O\left(u^4\right).\nonumber
\end{eqnarray}
\par 
The multidimensional pdf, (\ref{eq:p-Ising}), shows that we are essentially dealing 
with a beta pdf
\[p_{m}(u)=K(1+u)^{\half(m-1)}(1-u)^{\half(m-1)}+O\left(u^4\right),\]
defined over $[-1,+1]$, rather than over $[-1,0]$. In fact, the entropy reduction, 
(\ref{eq:Ising}) can be written as an entropy of mixing
\[
\Delta S(u)=-\half\left\{(1+u)\log(1+u)+(1-u)\log(1-u)\right\}. \]
\par
 The distribution of mass  inside a spherical
layer whose radius varies from $(1-\epsilon)$ to $1$, with $\epsilon$ an 
infinitesimal quantity is \cite[p. 191]{Finetti}
\begin{eqnarray*}
p_{m}(u)& = & K\left\{(1-u^2)^{\half(m-1)}-[(1-\epsilon)^2-u^2]^{\half(m-1)}\right\}
\\
& \simeq & K(m-1)\epsilon(1-u^2)^{\half(m-3)}=K(1-u^2)^{\half(m-3)},
\end{eqnarray*}
where the constants have been absorbed into the new definition of 
$K$. This is the same pdf as (\ref{eq:p-Ising}), except that $m$ has been reduced by
a factor of $2$. A change of variable, $u=\cos\theta$, converts 
it into
\begin{equation}
p_{m}(\theta)=K \sin^{m-2}\theta. \label{eq:theta}
\end{equation}
For a sphere, $m=3$, and it is well-known that two points on its surface subtend
an angle between $\theta$ and $\theta+d\theta$, at the center of the sphere, if and only
if the second point falls in a ring of area $2\pi\sin\theta\,d\theta$. This occurs with 
probability $2\pi\sin\theta\,d\theta/4\pi=\half\sin\theta\,d\theta$ \cite{Solomon}.
\par The approach to the normal pdf can be viewed in the following terms. 
As $m$ increases, jumps to the left and right will tend to equalize, like heads 
and tails, so that the distribution will tend to concentrate around the origin. 
To avoid this, and derive the shape of the asymptotic distribution, we 
introduce the scaling $u\rightarrow u/\sqrt{m}$ \cite[p. 59]{Finetti} so
that $\left(1-u^2/m\right)^{\half(m-1)}\rightarrow e^{-\half u^2}$.
The normal distribution is obtained in the limit
without having to invoke any assumptions regarding the central limit theorem.
\par The form of the pdf (\ref{eq:p-Ising}) agrees with that found from the 
log-likelihood equation for the centering parameter $\lambda$ \cite[p. 210]{Jeffreys}
\[\frac{d}{d\lambda}\log\mathcal
{L}(\lambda)=\sumn\frac{x_i-\lambda}{1-(x_i-\lambda)^2/m}.\]
Integrating gives the likelihood function 
\[\mathcal{L}(\lambda)=
\prod_{i=1}^n\left\{1-\frac{(x_i-\lambda)^2}{m}\right\}^{\half m},\]
of a Type II law. Apart from the norming constant, the likelihood function is seen
to be the product of pdfs of the form (\ref{eq:p-Ising}).
\par 
Analytic continuation of the exchange 
energy, (\emph{i.e.}, $J\rightarrow iJ$) converts the thermal equation of state 
(\ref{eq:beta}) into\footnote{Feller \cite{Feller} gives a $2$-dimensional example  where the
random equation of state, (\ref{eq:beta-bis}), for $u$ has the density 
(\ref{eq:Student}) with $m=2$. It consists of a light source at the origin on a wall
and a rotating mirror whose axis of rotation is at a unit distance from the wall. 
The angle of rotation is distributed uniformly between 
$-\half\pi$ and $\half\pi$, according to 
(\ref{eq:beta-bis}), and $u$ represents the distance between the light source and the point where the reflected ray
intersects the wall. The Cauchy pdf represents the distribution of the intensity of 
light along lines of constant $v$, the axis 
 orthogonal to $u$ in the $u,v$-plane. The convolution property of the Cauchy pdf, according to 
Feller, is a
statement of Hughen's principle which states that intensity of light is the same
for a source located at the origin along the line $v=1$, as it would be if the source
were distributed along this line and the intensity were measured along $v=2$. Moreover, if we make 
$n$ measurements, the average, $(U_1+U_2+\ldots+U_n)/n$, has exactly the same pdf
as that of a single measurement, so that nothing is gained by making repeated 
measurements.}
 \begin{equation}
\beta=\tan^{-1}u, \label{eq:beta-bis}
\end{equation}
where  $\beta\rightarrow i\beta$,  and $u\rightarrow -iu$ in 
(\ref{eq:beta}) 
\footnote{A similar transformation
leads from aperiodic to periodic behavior in Newton's reiteration scheme. An approximate
root $x_{n}$ of the equation $g(x)=0$ has $x_{n+1}=x_n-g(x_n)/g^{\prime}(x_n)$ as its
next approximation. For $g(x)=x^2-1$, this reads $x_{n+1}=\half(x_n+1/x_n)$. Setting
$x_n=\coth\theta_n$ gives $\theta_{n+1}=2\theta_n$, without any trace of
periodicity. However, by making $x$ purely imaginary, $x=iy$, the iterative scheme
$y_{n+1}=\half(y_n-1/y_n)$ has the solution $y=-\cot(\pi\vartheta)$, which demands that
$\vartheta_{n+1}=2\vartheta_n\mod 1$. Now, initial aperiodic behavior can lead to 
preperiodic orbits by an appropriate choice of a rational number for $\vartheta_n$, 
instead of to a fixed point \cite{Schroeder}.}. 
What was an inverse temperature has
now become an angle, $\beta$. Such a process  simulates a symmetric walk on an infinite 
$1$-dimensional lattice, where what was the dimensionless internal energy per
unit spin, $u$, is now the position of the walker along the chain. 
Since the interval for $\beta$ is $[-\half\pi,\half\pi]$, the walker can wander off 
to $\pm\infty$.
\par On account of the properties of concavity and convexity, the equation of state
(\ref{eq:beta-bis}) is the negative derivative of the entropy and the positive
derivative of the logarithm of the generating function, respectively. Differentiating a 
second time  shows that we are dealing with a Type VII distribution, whose weight 
is given by 
$\sigma_{+}$ in (\ref{eq:sigma}).
Integrating (\ref{eq:beta-bis}) gives the entropy reduction
\begin{equation}
\Delta S(u)=-u\tan^{-1}u+
\half\log\left(1+u^2\right)+\mbox{const.} 
\label{eq:S}
\end{equation}
For small $u$, the entropy reduction (\ref{eq:S}) reduces to the negative
quadratic form of the normal law, 
\[\Delta S(u)\simeq-\half\log\left(1+u^2\right)\rightarrow-\half u^2.\]\par
Introducing (\ref{eq:beta-bis}) into (\ref{eq:S}) enables the comparison of
\begin{equation}
\Delta S=-\beta\tan\beta-\log\cos\beta+\mbox{const.},
\label{eq:S-bis}
\end{equation}
and (\ref{eq:S-Ising}). Under $\beta\rightarrow i\beta$, one becomes the negative of
the other. On the strength of the Legendre transform, (\ref{eq:Legendre}), the
logarithm of the generating function can be identified from (\ref{eq:S-bis}) as
\begin{equation}\log\mathcal{Z}(\beta)=-\log\cos\beta.\label{eq:Z}
\end{equation}
\par The characteristic function  of a random
walk model with equal \emph{a priori\/} probabilities of a jump to the left or to the 
right is $\cos\beta$ \footnote{The generating function exists for all real $\beta$
in all cases where the pdf has a finite range, and in the case of the normal law
\cite[p. 87]{Jeffreys}. In cases where the integral exists, the logarithm of the 
generating function will be equal in magnitude and opposite in 
sign to the logarithm of the characteristic function.}. This symmetric Bernoullian distribution has mass $+\half$ at sites $+1$ and $-1$.
Since the jumps are iid, the characteristic function of $m$ jumps is simply
$\cos^m\beta$. The inverse Fourier transform
\begin{eqnarray*}p(u) & = &
\frac{2^{-m}}{\pi}\int_{-\half\pi}^{\half\pi}\,e^{-i\beta u}\left(e^{i\beta}
+e^{-i\beta}\right)^m\,d\beta\\
& = & \frac{1}{2^m}\left(\begin{array}{cc}
m\\ 
\half(m+u) \end{array}\right)\rightarrow e^{-u^2/2m}
\end{eqnarray*}
gives the pdf, which in the large $m$-limit, 
transforms into the normal pdf. The approach
to the normal pdf can be obtained from  Boltzmann's principle, 
(\ref{eq:Boltzmann}).
\par
With the entropy reduction given by (\ref{eq:S}), the pdf for a single 
jump of length $u$ is given by Boltzmann's principle, 
(\ref{eq:Boltzmann}):
\[
p(u)=K\left(1+u^2\right)^{\half}e^{-u
\tan^{-1}u}. \]
Expanding the exponent, we get
\begin{equation}
p(u)=K(1+u^2)^{-\half}+ O\left(u^4\right).\label{eq:Boltzmann-tris}
\end{equation}
If the steps, $U_1,U_2,\ldots,U_m$,  are iid, then the length,
\[u=||U||=\left(u_1^2+\cdots+u_m^2\right)^{\half},\]
will have an $m$-dimensional  pdf of the form \footnote{The distinction between
(\ref{eq:Student}) and the Cauchy distribution in $\Re^m$ is the following. For
$m=2$, the probability that a vector in a randomly chosen direction with a length
$R$ will be greater than $r$ is $\Pr(R>r)=(1+r^2)^{-\half}$ \cite[p. 71]{Feller}. 
The extreme observations
that $m-1$ will be greater than $r$ while the remaining one will lie in the range $dr$
will have a density proportional to $r(1+r^2)^{-\half(m+1)}$. The change of variable,
$r=\sqrt{x}$ will convert it into a beta density of the second kind, proportional
to $(1+x)^{-(\half m+1)}$. Likewise, the property of extreme observations for
$\Pr(R>r)=(1-r^2)^{\half}$ will lead to a beta density of the first kind, which
is closely allied to order statistics. For extreme
observations of largest value, Boltzmann's principle (\ref{eq:Boltzmann}) must be 
modified in such a way that the logarithm of the tail of the distribution is 
proportional to the entropy reduction \cite{Lavenda95}.}
\begin{equation}
p_m(u)=K
\left(1+u^2\right)^{-\half m}+O(u^4). 
\label{eq:Student}
\end{equation}
The change of variable, $u=\cot\theta$ will transform (\ref{eq:Student}) into 
(\ref{eq:theta}) \cite[p. 193]{Finetti}.\par
The form of the pdf (\ref{eq:Student}) is, again, supported by the log-likelihood equation
for the centering parameter $\lambda$ \cite[p. 210]{Jeffreys}
\[\frac{d}{d\lambda}\log\mathcal{L}(\lambda)=
\sumn\frac{x_i-\lambda}{1+(x_i-\lambda)^2/m}.\]
By integration we find
\[
\mathcal{L}(\lambda)=\prod_{i=1}^n\left\{1+\frac{(x_i-\lambda)^2}{m}\right\}^
{-\half m},\]
for the likelihood function of a Type VII law. To within a constant factor, the 
likelihood function is a product of pdfs of the form (\ref{eq:Student}).\par
The pdf (\ref{eq:Student}) is a generalization of the $1$-dimensional 
Student $t$-pdf with $m$ degrees of freedom to an $m$-dimensional distribution having 
the same number of degrees of freedom \cite[p. 194]{Finetti}. In the limit as 
$m\rightarrow\infty$,  (\ref{eq:Student}) transforms into the normal pdf,
\begin{equation}p(u)=\frac{1}{\sqrt{2\pi}}e^{-\half u^2},\label{eq:normal}
\end{equation}
where the norming constant $K$
\[\lim_{m\rightarrow\infty}\left\{m^{-\half}
\left[B\left(\half,\half(m-1)\right)\right]^{-1}\right\}=(2\pi)^{-\half},\]
and the scaling $u\rightarrow u/\sqrt{m}$ has been introduced. Student's 
$t$-distribution (\ref{eq:Student}) can, therefore,
rightly be considered the last stop before normality for large, but finite, $m$.\par
We have derived Student's $t$-pdf from  Boltzmann's principle,
(\ref{eq:Boltzmann}), where the entropy 
(\ref{eq:S}) is \emph{independent\/} of the number of degrees of freedom. 
As a consequence of their statistical independence, the entropy  is 
\emph{additive\/}. 
\emph{For $m$ idd random variables, or degrees of freedom, the total entropy is $m$ 
times as great as the entropy of a single variate, while the
joint pdf is the product of functions, and this can only be true if there is a 
logarithmic relation between the two\/.} If this
were not the case, it would be impossible to talk about the entropy of a single constituent
of the system, or the probability of a single event. Consequently, any putative expression 
for the entropy which is defined
in terms of the degrees of freedom of the system \cite{Tsallis} must be considered 
not as an entropy, but, rather, as an interpolation formula between two bona fide 
expressions of the entropy. A prototype is the R\'enyi entropy, which is an interpolation formula 
connecting the Boltzmann-Hartley and the  Gibbs-Shannon entropies as a parameter 
ranges over its permitted values (\textit{i.e.\/}, those that preserve the property
of Schur-concavity \cite{Lavenda98}).\par

\end{document}